\documentclass{article}

\usepackage{arxiv}

\usepackage[utf8]{inputenc} % allow utf-8 input
\usepackage[T1]{fontenc}    % use 8-bit T1 fonts
\usepackage{hyperref}       % hyperlinks
\usepackage{url}            % simple URL typesetting
\usepackage{booktabs}       % professional-quality tables
\usepackage{amsfonts}       % blackboard math symbols
\usepackage{nicefrac}       % compact symbols for 1/2, etc.
\usepackage{microtype}      % microtypography
\usepackage{lipsum}		
\usepackage{graphicx}
\usepackage{doi}

\usepackage{amsmath}
\usepackage{amssymb}
\newcommand{\norm}[1]{\left\lVert#1\right\rVert}
\usepackage[dvipsnames]{xcolor}
\usepackage{accents}
\usepackage{subcaption}
\allowdisplaybreaks

\title{An Adaptive Pilot Model with Reaction Time-Delay}

\date{} 				

\author{ \href{https://orcid.org/0000-0001-8598-6419}{\includegraphics[scale=0.06]{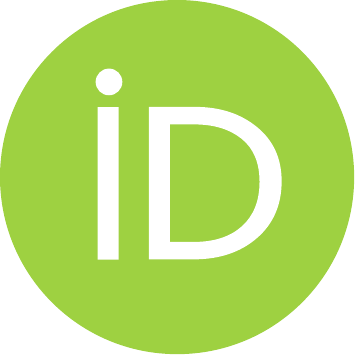}\hspace{1mm}Abdullah Habboush}\\
	Department of Mechanical Engineering\\
	Bilkent University\\
	Ankara 06800 \\
	\texttt{a.habboush@bilkent.edu.tr} \\
	\And
	\href{https://orcid.org/0000-0001-6270-5354}{\includegraphics[scale=0.06]{orcid.pdf}\hspace{1mm}Yildiray Yildiz} \\
	Department of Mechanical Engineering\\
	Bilkent University\\
	Ankara 06800 \\
	\texttt{yyildiz@bilkent.edu.tr} \\
}

\renewcommand{\shorttitle}{An Adaptive Pilot Model with Reaction Time-Delay}

\hypersetup{
	pdftitle={An Adaptive Pilot Model with Reaction Time-Delay},
	pdfauthor={Abdullah Habboush, Yildiray Yildiz},
	pdfkeywords={Adaptive control, human-in-the-loop control},
}

\begin{document}
	\maketitle
	
	\begin{abstract}
		Practical adaptive control implementations where human pilots coexist in the loop are still uncommon, despite their success in handling uncertain dynamical systems. This is owing to their special nonlinear characteristics which lead to unfavorable interactions between pilots and adaptive controllers. To pave the way for the implementation of adaptive controllers in piloted applications, we propose an adaptive human pilot model that takes into account the time delay in the pilot's response while operating on an adaptive control system. The model can be utilized in the evaluation of adaptive controllers through the simulation environment and guide in their design.
	\end{abstract}

	\keywords{Adaptive control \and human-in-the-loop control}

	\section{Introduction}
	Adaptive controllers are one of the major advancements in the field of control theory when it comes to addressing the control of dynamical systems that are prone to failures and uncertainties. While their design techniques are simple, and the theory behind their stability and performance is well established, their wide-spread use in real-life applications, where human pilots are in the loop, is yet to be seen. Several flight tests showed unfavorable interactions between human pilots and adaptive controllers due to their special nonlinear characteristics \cite{Klyde}. Hence, to aid in the design of adaptive controllers for piloted applications, a human-in-the-loop analysis is deemed necessary.
	
	Human pilot models play a crucial role in the evaluation of human-in-the-loop control systems as they allow the designer to test a controller through a simulation environment. Prominent models such as McRuer's crossover model \cite{McRuer} and its extensions \cite{Cooke,Rev2}, provide a simple fixed representation of human pilots in the loop with time-invariant control systems. However, such models fail to capture the adaptive behavior of human pilots when faced with unexpected anomalies. 
	
	A few adaptive human pilot models have been proposed in the literature. In \cite{Hess1} and \cite{Hess2}, an adaptive pilot model is proposed, where the adaptation laws are based on expert knowledge, aiming to make the adaptive pilot model follow the dictates of the crossover model. Inspired by this idea, an experimentally-validated adaptive pilot model is recently developed in \cite{Shahabpilot} and \cite{shahabpilot2}, by resorting to model reference adaptive control (MRAC) techniques which allows a rigorous stability analysis using the Lyapunov-Krasovskii stability criteria. These models assume that the pilot is operating on a linear control system, making them unsuitable for the evaluation of adaptive control systems. 
	Although there exist studies such as \cite{Tansel} and \cite{tohidi}, where an adaptive controller is in the loop, the pilot model used is not adaptive. 
	
	An adaptive human pilot model that is used in the loop with an adaptive controller has been recently proposed in \cite{Habboush} and \cite{habboushLCSS}. The development of the model is carried out based on MRAC architecture, with a rigorous Lyapunov stability analysis. The model does not explicitly take into account the time delay in human pilot's response, which narrows down the class of suitable applications.
	
	In this paper, we build upon the works in \cite{Habboush} and \cite{habboushLCSS}, by proposing an adaptive pilot model that considers the human internal time delay while operating on an adaptive control system. The model can be used for the evaluation of adaptive controllers in piloted applications and aid in their design. The inclusion of time delay in the human's response forms a major difficulty, which necessitates the prediction of the future states of a time-varying uncertain adaptive control system. We propose a novel approach by resorting to the fundamental theory of linear systems, and MRAC to provide a rigorous Lyapunov-Krasovskii stability analysis.
	
	The notation used here is standard, where $\mathbb{R}^{p\times q}$ [$\mathbb{S}^{p\times q}$]\{$\mathbb{D}^{p\times q}$\} denotes the set of real [symmetric real]\{diagonal real\} p by q matrices, and $\norm{.}$ refers to the euclidean norm for vectors ($q=1$), and the induced-2 norm for matrices. $\norm{.}_F$ refers to the Frobenius norm for matrices, $\mathrm{Tr\{.\}}$ refers to the trace operator, and $(.)^T$[$(.)^{-1}$] denotes the transpose [inverse] operator. $\text{Proj}(\hat{\theta}(t),Y)$ is the element-wise projection operator, defined in \cite{ShahabAuto}, used to bound each element $\hat{\theta}_{i,j}(t)$ of an adaptive parameter $\hat{\theta}(t)$ in a compact set $[\hat{\theta}_{min_{i,j}},\hat{\theta}_{max_{i,j}}]$. Finally, we write $\lambda_{min}(A)$ for the minimum eigenvalue of the matrix $A$ and we denote the set of positive definite real matrices by $\mathbb{R}_+^{p\times p}$.

	%%%%%%%%%%%%%%%%%%%%%%%%%%%%%%%%%%%%%%%%%%%%%%%%%%%%%%%%%%%%%%%%%%%%%%%%%%%%%%%%
	
	\section{Problem Statement}\label{problem statement}
	\begin{figure}[t]
		\centering
		\includegraphics[width=\linewidth]{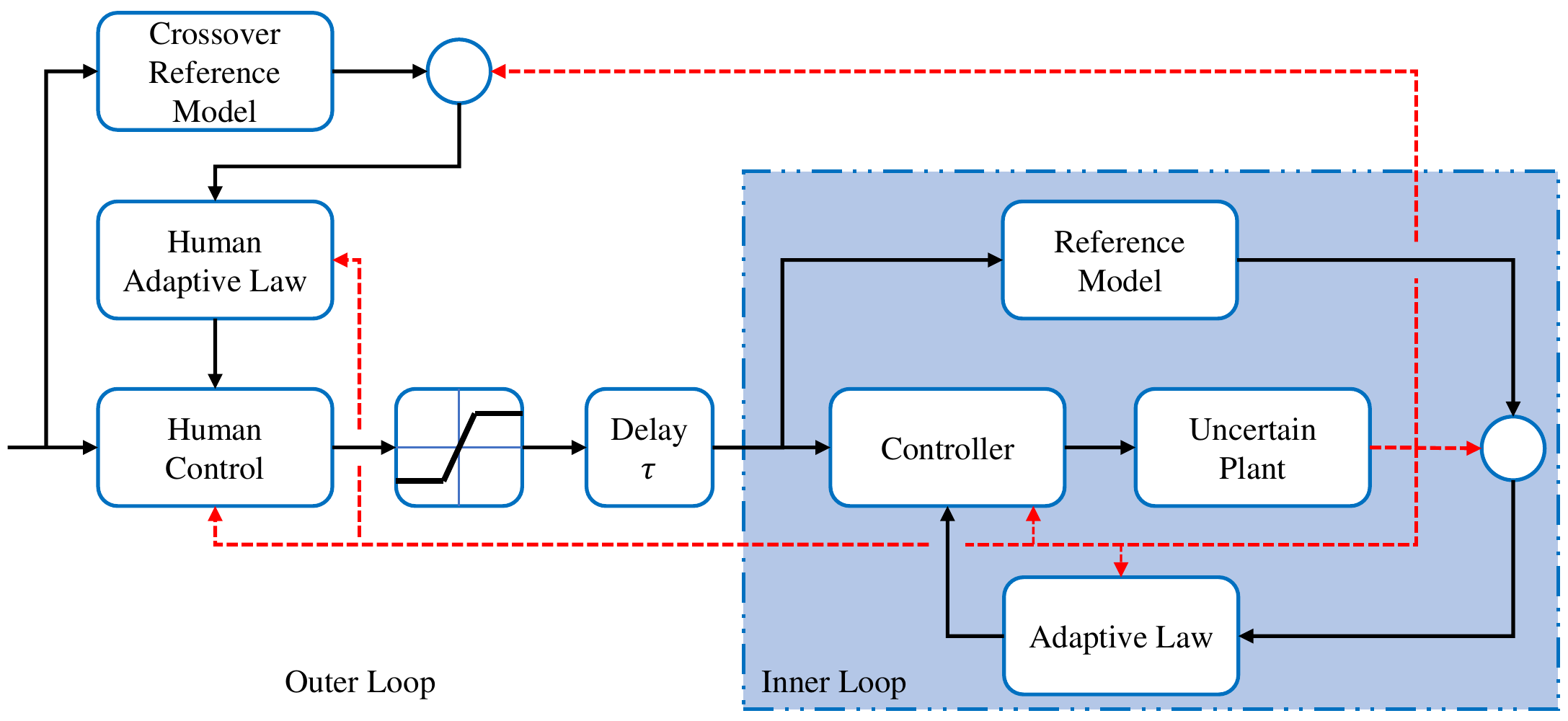}
		\caption{Block Diagram}
		\label{Block Diagram}
		\vspace{-4 mm}
	\end{figure}
	To model the human's adaptive control behavior with an adaptive controller in the loop, we start with a block diagram given in Fig. \ref{Block Diagram}. In the figure, the block diagram is divided into inner and outer loops. The inner-loop consists of an adaptive controller controlling a plant with uncertain dynamics such that the plant states follow those of a reference model by adjusting the control parameters using an adaptive law.
	
	The outer-loop consists of the human controlling the inner-loop such that the plant output follows a reference input. The human is assumed to be well trained, i.e, familiar with the nominal plant-controller dynamics. However, he/she is not aware of the uncertainties in the plant dynamics. This motivates modeling the human as an adaptive outer-loop controller, where an adaptive law is utilized to force the plant states to follow the states of the crossover-reference model.
	
	\section{Inner loop}\label{Inner loop}
	\noindent Consider the following uncertain plant dynamics
	\begin{equation} \label{plant dynamics}
		\begin{aligned}
			&\dot{x}_p (t)=A_px_p(t)+B_p \Lambda u_p(t),\\
			&y_1(t)=C_1^Tx_p(t),\,\,\,y_2(t)=C_2^Tx_p(t),
		\end{aligned}
	\end{equation}
	where $x_p(t)\in \mathbb{R}^{n_p}$ is the accessible state vector, $u_p(t)\in \mathbb{R}^m$ is the plant control input, $\Lambda \in \mathbb{R}^{m\times m}_+ \cap \mathbb{D}^{m\times m}$ is an unknown control effectiveness matrix with the diagonal elements $\lambda_{i,i}\in(0,1]$, $A_p \in \mathbb{R}^{n_p\times n_p}$ is an unknown system matrix, $B_p \in \mathbb{R}^{n_p\times m}$ is a known control input matrix, and $C_1\in \mathbb{R}^{n_p\times m}$ and $C_2\in \mathbb{R}^{n_p\times m}$ are both known output matrices. The outputs $y_1(t)\in \mathbb{R}^m$ and $y_2(t)\in \mathbb{R}^m$ are the outputs of interest for the inner and outer loops, respectively. Furthermore, it is assumed that the pair $(A_p, B_p)$ is controllable.
	
	Let the nominal plant dynamics be given as
	\begin{equation} \label{nominal plant}
		\dot{x}_n(t)=A_nx_n(t)+B_pu_n(t),
	\end{equation}
	where $u_n(t)\in \mathbb{R}^{m}$ is a nominal controller given as
	\begin{equation} \label{nominal controller}
		u_n(t)=-L_xx_n(t)+L_ry_h(t-\tau),
	\end{equation}
	where $y_h(t-\tau)\in \mathbb{R}^m$ is the human command to the inner-loop with an internal human time delay $\tau\in\mathbb{R}^+$, and $L_x\in \mathbb{R}^{m\times n_p}$ is such that $A_r\triangleq A_n-B_pL_x$ is Hurwitz. It is noted that the human input $y_h(t)$ is bounded due to physical manipulator limits. In the design of the outer loop, given in the following section, human input saturation bounds imposed by the manipulator limits are considered in the stability analysis. Defining $B_r \triangleq B_pL_r$, the reference model is assigned as
	\begin{equation} \label{reference model}
		\dot{x}_r(t)=A_rx_r(t)+B_ry_h(t-\tau),\,\, x_r(t_0)=0.
	\end{equation}
	For a constant $y_h$, at steady state, it is obtained using (\ref{reference model}) that $\dot{x}_r(\infty)=0=A_rx_r(\infty)+B_ry_h$, and therefore $x_r(\infty)=-A_r^{-1}B_pL_ry_h.$ This means that once the reference model state tracking is achieved, i.e., $\lim_{t\to \infty}x_p(t)=x_r(t)$, the plant output $y_1(t)$, given in (\ref{plant dynamics}), takes the form $y_1(\infty)=-C_1^TA_r^{-1}B_pL_ry_h.$
	To achieve $\lim_{t\to \infty}y_1(t)=y_h$, we select
	\begin{equation} \label{Lr}
		L_r=-(C_1^TA_r^{-1}B_p)^{-1}.
	\end{equation}
	Considering (\ref{plant dynamics}), we assume that there exist $K_x^*\in \mathbb{R}^{m\times n_p}$ and $K_r^*\in \mathbb{R}^{m\times m}$ such that the matching conditions
	\begin{equation} \label{matching conditions}
		\begin{aligned}
			A_p-B_p\Lambda K_x^*&=A_r\\
			B_p\Lambda K_r^*&=B_r\triangleq B_pL_r
		\end{aligned}
	\end{equation}
	are satisfied, where the second matching condition implies that $K_r^*=\Lambda^{-1}L_r$. We define the plant control law as 
	\begin{equation}\label{plant input}
		u_p(t)=-\hat{K}_x(t)x_p(t)+\text{diag}(\hat{\lambda}(t) )L_ry_h(t-\tau),
	\end{equation}
	where $\hat{K}_x(t)\in \mathbb{R}^{m\times n_p}$ and $\hat{\lambda}(t)\in \mathbb{R}^m$ are adjustable adaptive parameters serving as estimates for the ideal values $K_x^*$ and $\lambda^*$, respectively. It is noted that $\text{diag}\left(\lambda^* \right)=\Lambda^{-1}$ exists since $\Lambda$ is diagonal positive definite.
	
	Substituting (\ref{plant input}) into (\ref{plant dynamics}), one can rewrite (\ref{plant dynamics}) as
	\begin{equation} \label{plant dynamics2}
		\begin{aligned}
			\dot{x}_p(t)=A_rx_p(t)+B_ry_h(t-\tau)+B_p\Lambda\text{diag}(\tilde{\lambda}(t))L_ry_h(t-\tau)-B_p\Lambda\tilde{K}_x(t)x_p(t),
		\end{aligned}
	\end{equation}
	where $\tilde{K}_x(t)\triangleq \hat{K}_x(t)-K_x^*$ and $\tilde{\lambda}(t)\triangleq \hat{\lambda}(t)-\lambda^*$.
	
	By subtracting (\ref{reference model}) from (\ref{plant dynamics2}), and using $\Lambda\text{diag}(\tilde{\lambda}(t))L_ry_h(t-\tau)=\text{diag}(L_ry_h(t-\tau) )\Lambda \tilde{\lambda}(t) $, we obtain that
	\begin{equation}\label{error1}
		\dot{e}_1(t)=A_re_1(t)+B_p\text{diag}(L_ry_h(t-\tau) )\Lambda \tilde{\lambda}(t)-B_p\Lambda\tilde{K}_x(t)x_p(t),
	\end{equation}
	where $e_1(t)\triangleq x_p(t)-x_r(t)$ is the inner-loop tracking error. We define the inner-loop adaptive laws as
	\begin{subequations}\label{update laws}
		\begin{equation}
			\dot{\tilde{K}}_x^T(t)=\dot{\hat{K}}_x^T(t)=\gamma_x x_p(t)e_1(t)^TP_1B_p,
		\end{equation}
		\begin{equation}\label{lmbda}
			\dot{\tilde{\lambda}}(t)=\dot{\hat{\lambda}}(t)=\gamma_\lambda\text{Proj}\left(\hat{\lambda}(t),\: -\text{diag}(L_ry_h(t-\tau))B_p^TP_1e_1(t) \right), 
		\end{equation}
	\end{subequations}
	where positive bounds are set, by the projection operator, on each element $\hat{\lambda}_i(t)$, i.e., $\hat{\lambda}_{max_i}>\hat{\lambda}_{min_i}>0$ for all $i=1,\dots,m$. Furthermore, $\gamma_x,\gamma_\lambda \in \mathbb{R}_+$ are learning rates, and $P_1 \in \mathbb{R}^{n_p\times n_p}_+\cap \mathbb{S}^{n_p\times n_p}$ is the solution of the Lyapunov equation $A^T_rP_1+P_1A_r=-Q_1,$ for some $Q_1 \in \mathbb{R}^{n_p\times n_p}_+\cap \mathbb{S}^{n_p\times n_p}$. In this paper, without loss of generality, all learning rates are taken as scalars, instead of diagonal positive definite matrices, for simplicity of notation.
	
	\textbf{Lemma 1:} Consider the uncertain dynamical system (\ref{plant dynamics}), the reference model (\ref{reference model}), and the feedback control law given by (\ref{plant input}) and (\ref{update laws}). The solution $(e_1(t),\tilde{K}_x(t), \tilde{\lambda}(t))$ is Lyapunov stable in the large. Furthermore, since the human command $y_h(t)$ is bounded, due to imposed saturation limits by the physical manipulators, $\lim_{t\to \infty}e_1(t)=0$ and $\dot{\tilde{K}}_x(t)$ and $\dot{\tilde{\lambda}}(t)$ remain bounded along with all the signals in the inner-loop.
	
	\textit{Proof:} The proof of Lemma 1 can be found in \cite{habboushLCSS}. \hfill $\blacksquare$
	
	%%%%%%%%%%%%%%%%%%%%%%%%%%%%%%%%%%%%%%%%%%%%%%%%%%%%%%%%%%%%%%%%%%%%%%%%%%%%%%%%%%%%%%%%%
	%%%%%%%%%%%%%%%%%%%%%%%%%%%%%%%%%%%%%%%%%%%%%%%%%%%%%%%%%%%%%%%%%%%%%%%%%%%%%%%%%%%%%%%%%
	\section{Outer Loop}\label{OUTER LOOP}
	\noindent Using (\ref{matching conditions}), and $\text{diag}(\lambda^*)=\Lambda^{-1}$, (\ref{plant dynamics2}) can be rewritten as
	\begin{equation} \label{inner loop1}
		\dot{x}_p(t)=A_rx_p(t)+B_p\Lambda\text{diag}(\hat{\lambda}(t))L_ry_h(t-\tau)-B_p\Lambda\tilde{K}_x(t)x_p(t).
	\end{equation}
	Since we assume that the human operator is familiar with the nominal dynamics (\ref{nominal plant}) and (\ref{nominal controller}), the only unknowns in (\ref{inner loop1}) are $\Lambda$, $\hat{\lambda}(t)$ and $\tilde{K}_x(t)$. Furthermore, it is assumed that the internal time delay $\tau$ is known by the human pilot.
	
	Defining the unknown time-varying parameters as
	\begin{equation}\label{Lambda_2(t)}
		H^T(t)\triangleq -\Lambda \tilde{K}_x(t),\:\: \Lambda_2(t)\triangleq \Lambda\text{diag}(\hat{\lambda}(t)),
	\end{equation}
	
	equation (\ref{inner loop1}) can be rewritten as
	\begin{equation} \label{inner loop}
		\dot{x}_p(t)=(A_r+B_pH^T(t))x_p(t)+B_p\Lambda_2(t)L_ry_h(t-\tau).
	\end{equation}

	It is noted that although (\ref{inner loop}) is a non-linear control system, it is viewed by the pilot as a linear-time-varying system whose state matrix is represented by $A(t)=A_r+B_pH^T(t)$. The goal of the human is to control the system such that the plant states follow that of a unity feedback reference model with an open loop crossover model transfer function. We refer to the latter as the \textit{crossover-reference model}  (Fig. \ref{Block Diagram}). Let the crossover-reference model be given as
	\begin{equation} \label{crossover refernece model}
		\dot{x}_m(t)=A_mx_m(t)+B_mr(t-\tau),
	\end{equation}
	where $x_m(t)\in \mathbb{R}^{n_p}$ is the crossover-reference model state vector, $r(t)\in \mathbb{R}^m$ is a bounded reference input, $A_m\in \mathbb{R}^{n_p\times n_p}$ is Hurwitz and $B_m\triangleq B_r\theta_r\in \mathbb{R}^{n_p\times m}$. Similar to the inner-loop, and for a constant reference input $r$, the nominal feed-forward gain $\theta_r\in \mathbb{R}^{m\times m}$ is selected as
	\begin{equation}\label{thr}
		\theta_r=-(C_2^TA_m^{-1}B_r)^{-1}
	\end{equation}
	to achieve $\lim_{t\to \infty}y_2(t)=r$ if $\lim_{t\to \infty}x_p(t)=x_m(t)$.
	
	In an ideal case where the human input is not saturated, and both $H(t)$ and $\Lambda_2(t)$ are known, the following non-causal control law achieves the crossover-reference model dynamics
	\begin{equation}\label{ideal}
		\begin{aligned}
			\mathcal{G}^*(t)&=-\theta_xx_p(t+\tau)+\theta_rr(t)-L_r^{-1}H^T(t+\tau)x_p(t+\tau),\\
			y_h^*(t)&=L_r^{-1}\Lambda_2^{-1}(t+\tau)L_r\mathcal{G}^*(t),
		\end{aligned}
	\end{equation}
	where we assume that there exists $\theta_x\in \mathbb{R}^{m\times n_p}$ such that $A_m= A_r-B_pL_r\theta_x$. The future state of the plant is predicted by solving the time-varying differential equation (\ref{inner loop}) as
	\begin{equation}\label{predictor}
		\begin{aligned}
			x_p(t+\tau)=\Phi(t+\tau,t)x_p(t)+\int_{-\tau}^{0}\Phi(t+\tau,t+\eta+\tau)B_p\Lambda_2(t+\eta+\tau)L_ry_h(t+\eta)\,\text{d}\eta,
		\end{aligned}
	\end{equation}
	where $\Phi(t_2,t_1)\in \mathbb{R}^{n_p\times n_p}$ is the state transition matrix of (\ref{inner loop}). Motivated by (\ref{ideal}) and (\ref{predictor}), we define the human control input as
	\begin{subequations} \label{human input}
		\begin{align}
			\mathcal{G}(t)&=\hat{\Phi}_1(t)x_p(t)+\theta_rr(t)+\int_{-\tau}^{0}\hat{\Phi}_2(t,\eta)L_ry_h(t+\eta)\,\text{d}\eta, \label{G}\\
			v(t)&=L_r^{-1}\text{diag}(\hat{\lambda}_{2}(t))L_r\mathcal{G}(t), \label{v}\\
			y_{h_i}(t)&= \label{saturation}
			\begin{cases}
				v_i(t),& \text{if}\; |v_i(t)|\leq y_{o_i},\\
				y_{o_i}\mathrm{sgn}(v_i(t)),& \text{if}\; |v_i(t)|> y_{o_i},
			\end{cases}
		\end{align}
	\end{subequations}
	where $\hat{\Phi}_1(t)\in \mathbb{R}^{m\times n_p}$, $\hat{\Phi}_2(t,\eta)\in \mathbb{R}^{m\times m}$ and $\hat{\lambda}_2(t)\in \mathbb{R}^m$ are adaptive parameters serving as estimates for the ideal values
	\begin{equation}\label{ideal values}
		\begin{aligned}
			\Phi_1^*(t)&=\bar{H}(t)\Phi(t+\tau,t),\\
			\Phi_2^*(t,\eta)&=\bar{H}(t)\Phi(t+\tau,t+\eta+\tau)B_p\Lambda_2(t+\eta+\tau)
		\end{aligned}
	\end{equation}
	and $\lambda_2^*(t)$, respectively, where $\bar{H}(t)\triangleq-(\theta_x+L_r^{-1}H^T(t+\tau))$. It is noted that $\text{diag}\left(\lambda^*_2(t) \right)=\Lambda^{-1}_2(t+\tau)$ exists for all $t\geq 0$ since $\Lambda_2(t)$, defined in (\ref{Lambda_2(t)}), is diagonal positive definite at all time instants. This is guaranteed due to the positive lower bounds imposed by the projection operator in (\ref{lmbda}) on the inner-loop adaptive parameter $\hat{\lambda}(t)$. Furthermore, (\ref{saturation}) is an element-wise saturation function where $y_{o_i}\in \mathbb{R}_+$ is the saturation limit of $y_{h_i}(t)$ (the $i^{\mathrm{th}}$ element of $y_h(t)$).
	
	Substituting (\ref{human input}) into (\ref{inner loop}), and with some algebraic manipulations, we obtain that
	\begin{equation} \label{system2}
		\begin{aligned}
			\dot{x}_p(t)=&A_mx_p(t)+B_mr(t-\tau)+B_p\Lambda_2(t)L_r\Delta y(t-\tau)+B_p\Lambda_2(t)\text{diag}(\tilde{\lambda}_2(t-\tau))L_r\mathcal{G}(t-\tau)\\
			&+B_pL_r\tilde{\Phi}_1(t-\tau)x_p(t-\tau)+B_pL_r\int_{-\tau}^{0}\tilde{\Phi}_2(t-\tau,\eta)L_ry_h(t+\eta-\tau)\text{d}\eta,
		\end{aligned}
	\end{equation}
	where $\tilde{\Phi}_1(t)\triangleq \hat{\Phi}_1(t)-\Phi_1^*(t)$, $\tilde{\Phi}_2(t,\eta)\triangleq \hat{\Phi}_2(t,\eta)-\Phi_2^*(t,\eta)$ and $\tilde{\lambda}_2(t)\triangleq \hat{\lambda}_2(t)-\lambda_2^*(t)$ are outer-loop adaptive parameters errors, and $\Delta y(t)\triangleq y_h(t)-v(t)$ is the control deficiency due to human input saturation. 
	
	Subtracting (\ref{crossover refernece model}) from (\ref{system2}), and using $\Lambda_2\text{diag}(\tilde{\lambda}_2)L_r\mathcal{G}=\text{diag}(L_r\mathcal{G})\Lambda_2 \tilde{\lambda}_2 $ results in the outer-loop error dynamics 
	\begin{equation}\label{error2}
		\begin{aligned}
			\dot{e}_2(t)=&A_me_2(t)+B_p\Lambda_2(t)L_r\Delta y(t-\tau)+B_p\text{diag}(L_r\mathcal{G}(t-\tau))\Lambda_2(t)\tilde{\lambda}_2(t-\tau)\\
			&+B_pL_r\tilde{\Phi}_1(t-\tau)x_p(t-\tau)+B_pL_r\int_{-\tau}^{0}\tilde{\Phi}_2(t-\tau,\eta)L_ry_h(t+\eta-\tau)\text{d}\eta,
		\end{aligned}
	\end{equation}
	where $e_2(t)\triangleq x_p(t)-x_m(t)$ is the outer-loop tracking error.
	
	We generate an auxiliary signal $e_\Delta (t)$ as in \cite{Annaswamy,Schwager}
	\begin{equation}\label{auxilliary error}
		\begin{aligned}
			&\dot{e}_\Delta(t)=A_me_\Delta(t)+B_p\text{diag}(\hat{\lambda}_3(t))L_r\Delta y(t-\tau),\\
			&e_\Delta(t_0)=0,
		\end{aligned}
	\end{equation}
	where $\hat{\lambda}_3(t)\in \mathbb{R}^m$ is an adjustable adaptive parameter serving as an estimate for the ideal value $\lambda_3^*(t)$, and $\text{diag}(\lambda_3^*(t))=\Lambda_2(t)$.
	Defining an augmented error signal as $e_y(t)\triangleq e_2(t)-e_\Delta(t)$, and exploiting the fact that $\text{diag}(\tilde{\lambda}_3)L_r\Delta y=\text{diag}(L_r\Delta y)\tilde{\lambda}_3$ yields
	\begin{equation} \label{augmented error}
		\begin{aligned}
			\dot{e}_y(t)=&A_me_y(t)-B_p\text{diag}(L_r\Delta y(t-\tau))\tilde{\lambda}_3(t)+B_p\text{diag}(L_r\mathcal{G}(t-\tau))\Lambda_2(t)\tilde{\lambda}_2(t-\tau)\\
			&+B_pL_r\tilde{\Phi}_1(t-\tau)x_p(t-\tau)+B_pL_r\int_{-\tau}^{0}\tilde{\Phi}_2(t-\tau,\eta)L_ry_h(t+\eta-\tau)\text{d}\eta,
		\end{aligned}
	\end{equation}
	where $\tilde{\lambda}_3(t)\triangleq \hat{\lambda}_3(t)-\lambda_3^*(t)$. Equation (\ref{augmented error}) is in a standard error model form \cite{yildizposi,Shahabpilot}. We propose the adaptive laws
	\begin{subequations}\label{outer update laws}
		\begin{align}
			&\dot{\hat{\lambda}}_2(t)=\gamma_2\text{Proj}\left( \hat{\lambda}_2(t),\,-\text{diag}(L_r\mathcal{G}(t-\tau))B_p^TP_2e_y(t)\right),\\
			&\dot{\hat{\lambda}}_3(t)=\gamma_3\text{Proj}\left(\hat{\lambda}_3(t), \, \text{diag}(L_r\Delta y(t-\tau))B_p^TP_2e_y(t) \right),\\
			&\dot{\hat{\Phi}}_1^T(t)=\gamma_{\phi_1}\text{Proj}\left(\hat{\Phi}_1^T(t),\,-x_p(t-\tau)e_y^T(t)P_2B_pL_r \right),\\
			&\dot{\hat{\Phi}}_2^T(t,\eta)=\gamma_{\phi_2}\text{Proj}\left(\hat{\Phi}_2^T(t,\eta),\, -L_ry_h(t+\eta-\tau)e_y^T(t)P_2B_pL_r \right), \label{phi2dot}
		\end{align}
	\end{subequations}
	where $\gamma_2,\gamma_3,\gamma_{\phi_1}, \gamma_{\phi_2} \in \mathbb{R}_+$ are learning rates, and $P_2 \in \mathbb{R}^{n_p\times n_p}_+\cap \mathbb{S}^{n_p\times n_p}$ is the solution of the Lyapunov equation $A^T_mP_2+P_2A_m=-Q_2,$ for some $Q_2 \in \mathbb{R}^{n_p\times n_p}_+\cap \mathbb{S}^{n_p\times n_p}$.
	
	The following Lemma establishes key bounds on the state transition matrix of (\ref{inner loop}) and its time derivative, which is then utilized in the remarks that follow to show that all ideal values, of the outer-loop adaptive parameters, and their time derivatives are bounded. Such bounds play a crucial role in the stability proof that follows in Theorem 1.
	
	\textbf{Lemma 2:} The state transition matrix $\Phi(t+\tau,t)$ and its time derivative $\dot{\Phi}(t+\tau,t)$ are bounded, i.e., there exist $\phi\in \mathbb{R}_+$ and $\dot{\phi}\in\mathbb{R}_+$ such that $\norm{\Phi(t+\tau,t)}_F\leq \phi$ and $\norm{\dot{\Phi}(t+\tau,t)}_F\leq \dot{\phi}$ for all $t\geq t_0$. In addition, the same bounds apply for $\Phi(t+\tau,t+\eta+\tau)$ and its time derivative for all $t\geq t_0$, $-\tau\leq\eta\leq0$.
	
	\textit{Proof:} It follows from Lemma 1 that the origin $\{e_1=0,\,\tilde{K}_x=0,\,\tilde{\lambda}=0\}$ is uniformly stable in the large. The state transition matrix $\Phi(t+\tau,t)$ defines the solution $x_p(t+\tau)=\Phi(t+\tau,t)x_p(t)$ of the homogeneous part of (\ref{inner loop})
	\begin{equation}\label{homogeneous inner loop}
		\dot{x}_p(t)=(A_r+B_pH^T(t))x_p(t).
	\end{equation}
	Since $x_p(t)=e_1(t)+x_r(t)$, and $x_r(t_0)=0$, then the homogeneous part of (\ref{reference model})
	\begin{equation}
		\dot{x}_r(t)=A_rx_r(t), \,\, x_r(t_0)=0,
	\end{equation}
	yields the solution $x_r(t)=0$, which implies that $x_p(t)=e_1(t)$ for all $t\geq t_0$. This shows that the origin $x_p=0$ of (\ref{homogeneous inner loop}) is uniformly stable in the large. It then follows from Theorem 6.4 in \cite{rugh} that there exists $\phi\in \mathbb{R}_+$ such that $\norm{\Phi(t_2,t_1)}_F\leq \phi$ for all $t_1,\, t_2$, where $t_2\geq t_1$. This in turn implies that
	\begin{equation}\label{bdd Phi}
		\begin{aligned}
			\norm{\Phi(t+\tau,t)}_F\leq\phi,\,\, &\text{for all} \,\, t\geq t_0,\\
			\norm{\Phi(t+\tau,t+\eta+\tau)}_F\leq\phi,\,\, &\text{for all} \,\, t\geq t_0, \,\, -\tau\leq\eta\leq0.
		\end{aligned}
	\end{equation}
	Furthermore, the state transition matrix can be defined by the so-called fundamental matrix $X(t)$ of (\ref{homogeneous inner loop}) as
	\begin{equation}\label{phi in terms of X}
		\Phi(t+\tau,t)=X(t+\tau)X(t)^{-1},
	\end{equation}
	where $X(t)$ is non-singular for all $t\geq t_0$ \cite{chen}, and satisfies $\dot{X}(t)=A(t)X(t)$. Denoting $A(t)\triangleq(A_r+B_pH^T(t))$, and differentiating (\ref{phi in terms of X}) yields 
	\begin{equation}\label{Phi dot}
		\begin{aligned}
			\dot{\Phi}(t+\tau,t)=&\dot{X}(t+\tau)X(t)^{-1}+X(t+\tau)\frac{\text{d}}{\text{d}t}X(t)^{-1}\\
			=&\dot{X}(t+\tau)X(t)^{-1}-X(t+\tau)X(t)^{-1}\dot{X}(t)X(t)^{-1}\\
			=&A(t+\tau)\Phi(t+\tau,t)-\Phi(t+\tau,t)A(t).
		\end{aligned}
	\end{equation}
	Since the boundedness of $H^T(t)\triangleq -\Lambda \tilde{K}_x(t)$ follows from Lemma 1, then $A(t)$ is also bounded. Together with (\ref{bdd Phi}), this shows that all the terms in (\ref{Phi dot}) are bounded, which implies that there exists $\dot{\phi}\in\mathbb{R}_+$ such that 
	\begin{equation} \label{bdd Phidot}
		\begin{aligned}
			\norm{\dot{\Phi}(t+\tau,t)}_F\leq \dot{\phi}, \,\, &\text{for all} \,\, t\geq t_0,\\
			\norm{\dot{\Phi}(t+\tau,t+\eta+\tau)}_F\leq\dot{\phi},\,\, &\text{for all} \,\, t\geq t_0, \,\, -\tau\leq\eta\leq0.
		\end{aligned}
	\end{equation}\hfill $\blacksquare$
	
	\textbf{Remark 1:} It follows from Lemma 1 that $\tilde{K}(t)$, $\hat{\lambda}(t)$, $\dot{\tilde{K}}(t)$ and $\dot{\hat{\lambda}}(t)$ are bounded, which implies the boundedness of ${H}(t)$, $\Lambda_2(t)$, $\dot{H}(t)$ and $\dot{\Lambda}_2(t)$. Therefore, there exist $h\in \mathbb{R}_+$, $\dot{h}\in\mathbb{R}_+$, $\beta_3\in\mathbb{R}_+$ and $\dot{\beta}_3\in\mathbb{R}_+$ such that $\norm{H(t)}\leq h$, $\norm{\dot{H}(t)}\leq \dot{h}$, $\norm{\Lambda_2(t)}_F\leq\beta_3$ and $\norm{\dot{\Lambda}_2(t)}_F\leq\dot{\beta}_3$ for all $t\geq t_0$. The latter implies that $\norm{\lambda_3^*(t)}\leq\beta_3$ and $\norm{\dot{\lambda}_3^*(t)}\leq\dot{\beta}_3$. Moreover, as $\hat{\lambda}_{min_i}>0$ for $i=1,\dots,m$, there exists $\beta_2\in \mathbb{R}_+$ such that $\norm{\Lambda_2^{-1}(t)}_F\leq \beta_2$. And since $\frac{\text{d}\Lambda_2^{-1}}{\text{d}t}=-\Lambda_2^{-1}\dot{\Lambda}_2\Lambda_2^{-1}$, then there exists $\dot{\beta}_2\in \mathbb{R}_+$ such that $\norm{\frac{{\text{d}\Lambda}_2^{-1}}{\text{d}t}}_F\leq \dot{\beta}_2$. This implies that $\norm{\lambda_2^*(t)}\leq\beta_2$ and $\norm{\dot{\lambda}_2^*(t)}\leq\dot{\beta}_2$ for all $t\geq t_0$.
	
	\textbf{Remark 2:} Together with Remark 1, the bounds (\ref{bdd Phi}) and (\ref{bdd Phidot}), established in Lemma 2, show that all the terms of the ideal values (\ref{ideal values}) and their time derivatives are bounded. Hence, there exist $\phi_1,\dot{\phi}_1,\phi_2,\dot{\phi}_2\in\mathbb{R}_+$ such that $\norm{\Phi_1^*(t)}_F\leq\phi_1$, $\norm{\dot{\Phi}_1^*(t)}_F\leq \dot{\phi}_1$ for all $t\geq t_0$, and $\norm{\Phi_2^*(t,\eta)}_F\leq\phi_2$, $\norm{\dot{\Phi}_2^*(t,\eta)}_F\leq \dot{\phi}_2$ for all $t\geq t_0$, $-\tau\leq\eta\leq0$.

	\textbf{Theorem 1:} Consider the uncertain dynamical system given by (\ref{plant dynamics}), the adaptive controller given by (\ref{reference model}), (\ref{plant input}) and (\ref{update laws}), and the adaptive human pilot model given by (\ref{crossover refernece model}), (\ref{human input}) and (\ref{outer update laws}). Then, there exists $\tau^*\in\mathbb{R}_+$ such that for all $\tau\in[0,\,\tau^*]$, the solution $(e_y(t), \tilde{\lambda}_2(t),\tilde{\lambda}_3(t),\tilde{\Phi}_1(t), \tilde{\Phi}_2(t,\eta))$ remains bounded for all $t\geq t_0$ and converges to the compact set defined in (\ref{E}). Furthermore, the closed-loop system is stable in the large, and all signals are bounded.
	
	\textit{Proof:} Consider the Lyapunov-Krasovskii functional
	\begin{equation}\label{V}
		\begin{aligned}
			V_2&=e_y^T(t)P_2e_y(t)+\gamma_3^{-1}\tilde{\lambda}_3^T(t)\tilde{\lambda}_3(t)\\
			&+\gamma_2^{-1}\tilde{\lambda}_2^T(t)\Lambda_2(t)\tilde{\lambda}_2(t)+\int_{-\tau}^{0}\int_{t+\nu}^{t}\dot{\tilde{\lambda}}_2^T(\xi)\dot{\tilde{\lambda}}_2(\xi)\text{d}\xi\text{d}\nu\\
			&+\gamma_{\phi_1}^{-1}\text{Tr}\{\tilde{\Phi}_1^T(t)\tilde{\Phi}_1(t)\}+\int_{-\tau}^{0}\int_{t+\nu}^{t}\text{Tr}\{\dot{\tilde{\Phi}}_1^T(\xi)\dot{\tilde{\Phi}}_1(\xi)\}\text{d}\xi\text{d}\nu\\
			&+\gamma_{\phi_2}^{-1}\int_{-\tau}^{0}\text{Tr}\{\tilde{\Phi}_2^T(t,\eta)\tilde{\Phi}_2(t,\eta)\}\text{d}\eta+\int_{-\tau}^{0}\int_{t+\nu}^{t}\int_{-\tau}^{0}\text{Tr}\{\dot{\tilde{\Phi}}_2^T(\xi,\eta)\dot{\tilde{\Phi}}_2(\xi,\eta)\}\text{d}\eta\text{d}\xi\text{d}\nu.
		\end{aligned}
	\end{equation}
	For brevity, we define 
	\begin{equation}
		\begin{aligned}
			W(t)&\triangleq
			\int_{-\tau}^{0}\int_{t+\nu}^{t}\dot{\tilde{\lambda}}_2^T(\xi)\dot{\tilde{\lambda}}_2(\xi)\text{d}\xi\text{d}\nu+\int_{-\tau}^{0}\int_{t+\nu}^{t}\text{Tr}\{\dot{\tilde{\Phi}}_1^T(\xi)\dot{\tilde{\Phi}}_1(\xi)\}\text{d}\xi\text{d}\nu\\
			&+\int_{-\tau}^{0}\int_{t+\nu}^{t}\int_{-\tau}^{0}\text{Tr}\{\dot{\tilde{\Phi}}_2^T(\xi,\eta)\dot{\tilde{\Phi}}_2(\xi,\eta)\}\text{d}\eta\text{d}\xi\text{d}\nu,
		\end{aligned}
	\end{equation}
	where
	\begin{equation}\label{Wdot}
		\begin{aligned}
			\dot{W}(t)&=
			\tau\dot{\tilde{\lambda}}_2^T(t)\dot{\tilde{\lambda}}_2(t)-\int_{-\tau}^{0}\dot{\tilde{\lambda}}_2^T(t+\nu)\dot{\tilde{\lambda}}_2(t+\nu)\text{d}\nu+\tau\text{Tr}\{\dot{\tilde{\Phi}}_1^T(t)\dot{\tilde{\Phi}}_1(t)\}-\int_{-\tau}^{0}\text{Tr}\{\dot{\tilde{\Phi}}_1^T(t+\nu)\dot{\tilde{\Phi}}_1(t+\nu)\}\text{d}\nu\\
			&+\tau\int_{-\tau}^{0}\text{Tr}\{\dot{\tilde{\Phi}}_2^T(t,\eta)\dot{\tilde{\Phi}}_2(t,\eta)\} \text{d}\eta-\int_{-\tau}^{0}\int_{-\tau}^{0}\text{Tr}\{\dot{\tilde{\Phi}}_2^T(t+\nu,\eta)\dot{\tilde{\Phi}}_2(t+\nu,\eta)\}\text{d}\eta\text{d}\nu.
		\end{aligned}
	\end{equation}
	Differentiating (\ref{V}) along the trajectories (\ref{augmented error}) and (\ref{outer update laws}), and using (\ref{Wdot}), we obtain that
	\begin{equation*}
		\begin{aligned}
			\dot{V}_2=&\dot{e}_y^T(t)P_2e_y(t)+e_y^T(t)P_2\dot{e}_y(t)+2\gamma_3^{-1}\tilde{\lambda}_3^T(t)\dot{\tilde{\lambda}}_3(t)\\
			&+2\gamma_2^{-1}\tilde{\lambda}_2^T(t)\Lambda_2(t)\dot{\tilde{\lambda}}_2(t)+\gamma_2^{-1}\tilde{\lambda}_2^T(t)\dot{\Lambda}_2(t)\tilde{\lambda}_2(t)\\
			&+2\gamma_{\phi_1}^{-1}\text{Tr}\{\dot{\tilde{\Phi}}_1^T(t)\tilde{\Phi}_1(t)\}+2\gamma_{\phi_2}^{-1}\int_{-\tau}^{0}\text{Tr}\{\dot{\tilde{\Phi}}_2^T(t,\eta)\tilde{\Phi}_2(t,\eta)\}\text{d}\eta+\dot{W}(t)\\
			=&-e_y^T(t)Q_2e_y(t)-2\tilde{\lambda}_3^T(t)\text{diag}(L_r\Delta y(t-\tau))B_p^TP_2e_y(t)\\
			&+2\tilde{\lambda}_2^T(t-\tau)\Lambda_2(t)\text{diag}(L_r\mathcal{G}(t-\tau))B_p^TP_2e_y(t)\\
			&+2e_y^T(t)P_2B_pL_r\tilde{\Phi}_1(t-\tau)x_p(t-\tau)\\
			&+2e_y^T(t)P_2B_pL_r\int_{-\tau}^{0}\tilde{\Phi}_2(t-\tau,\eta)L_ry_h(t+\eta-\tau)\text{d}\eta\\
			&+2\gamma_3^{-1}\tilde{\lambda}_3^T(t)\dot{\tilde{\lambda}}_3(t)+2\gamma_{\phi_1}^{-1}\text{Tr}\{\dot{\tilde{\Phi}}_1^T(t)\tilde{\Phi}_1(t)\}\\
			&+2\gamma_2^{-1}\tilde{\lambda}_2^T(t)\Lambda_2(t)\dot{\tilde{\lambda}}_2(t)+\gamma_2^{-1}\tilde{\lambda}_2^T(t)\dot{\Lambda}_2(t)\tilde{\lambda}_2(t)\\
			&+2\gamma_{\phi_2}^{-1}\int_{-\tau}^{0}\text{Tr}\{\dot{\tilde{\Phi}}_2^T(t,\eta)\tilde{\Phi}_2(t,\eta)\}\text{d}\eta+\dot{W}(t).\\
		\end{aligned}
	\end{equation*}
	Using the fact that $g(t-\tau)=g(t)-\int_{-\tau}^{0}\dot{g}(t+\nu)\text{d}\nu$ for $\tilde{\lambda}_2^T(t-\tau)$, $\tilde{\Phi}_1(t-\tau)$ and $\tilde{\Phi}_2(t-\tau,\eta)$, and decomposing $\dot{\tilde{\lambda}}_2(t)=\dot{\hat{\lambda}}_2(t)-\dot{\lambda}_2^*(t)$, $\dot{\tilde{\lambda}}_3(t)=\dot{\hat{\lambda}}_3(t)-\dot{\lambda}_3^*(t)$, $\dot{\tilde{\Phi}}_1(t)=\dot{\hat{\Phi}}_1(t)-\dot{\Phi}_1^*(t)$ and $\dot{\tilde{\Phi}}_2(t,\eta)=\dot{\hat{\Phi}}_2(t,\eta)-\dot{\Phi}_2^*(t,\eta)$, we get
	\begin{equation*}
		\begin{aligned}
			\dot{V}_2=&-e_y^T(t)Q_2e_y(t)-2\tilde{\lambda}_3^T(t)\text{diag}(L_r\Delta y(t-\tau))B_p^TP_2e_y(t)\\
			&+2\tilde{\lambda}_2^T(t)\Lambda_2(t)\text{diag}(L_r\mathcal{G}(t-\tau))B_p^TP_2e_y(t)\\
			&-2\left(\int_{-\tau}^{0}\dot{\tilde{\lambda}}_2^T(t+\nu)\text{d}\nu\right)\Lambda_2(t)\text{diag}(L_r\mathcal{G}(t-\tau))B_p^TP_2e_y(t)\\
			&+2e_y^T(t)P_2B_pL_r\bigg[\tilde{\Phi}_1(t)x_p(t-\tau)-\left(\int_{-\tau}^{0}\dot{\tilde{\Phi}}_1(t+\nu)\text{d}\nu\right)x_p(t-\tau)\\
			&+\int_{-\tau}^{0}\tilde{\Phi}_2(t,\eta)L_ry_h(t+\eta-\tau)\text{d}\eta-\int_{-\tau}^{0}\left(\int_{-\tau}^{0}\dot{\tilde{\Phi}}_2(t+\nu,\eta)\text{d}\nu\right)L_ry_h(t+\eta-\tau)\text{d}\eta\bigg]\\
			&+2\gamma_3^{-1}\tilde{\lambda}_3^T(t)\dot{\hat{\lambda}}_3(t)-2\gamma_3^{-1}\tilde{\lambda}_3^T(t)\dot{\lambda}^*_3(t)+2\gamma_2^{-1}\tilde{\lambda}_2^T(t)\Lambda_2(t)\dot{\hat{\lambda}}_2(t)-2\gamma_2^{-1}\tilde{\lambda}_2^T(t)\Lambda_2(t)\dot{\lambda}^*_2(t)\\
			&+\gamma_2^{-1}\tilde{\lambda}_2^T(t)\dot{\Lambda}_2(t)\tilde{\lambda}_2(t)+2\gamma_{\phi_1}^{-1}\text{Tr}\{\dot{\hat{\Phi}}_1^T(t)\tilde{\Phi}_1(t)\}-2\gamma_{\phi_1}^{-1}\text{Tr}\{\dot{\Phi}_1^{*T}(t)\tilde{\Phi}_1(t)\}\\
			&+2\gamma_{\phi_2}^{-1}\int_{-\tau}^{0}\text{Tr}\{\dot{\hat{\Phi}}_2^T(t,\eta)\tilde{\Phi}_2(t,\eta)\}\text{d}\eta-2\gamma_{\phi_2}^{-1}\int_{-\tau}^{0}\text{Tr}\{\dot{\Phi}_2^{*T}(t,\eta)\tilde{\Phi}_2(t,\eta)\}\text{d}\eta+\dot{W}(t).\\
		\end{aligned}
	\end{equation*}
	Defining 
	\begin{equation}\label{N*}
		\begin{aligned}
			N^*(t)&\triangleq \gamma_2^{-1}\tilde{\lambda}_2^T(t)\dot{\Lambda}_2(t)\tilde{\lambda}_2(t)-2\gamma_2^{-1}\tilde{\lambda}_2^T(t)\Lambda_2(t)\dot{\lambda}^*_2(t)\\
			&-2\gamma_3^{-1}\tilde{\lambda}_3^T(t)\dot{\lambda}^*_3(t)-2\gamma_{\phi_1}^{-1}\text{Tr}\{\dot{\Phi}_1^{*T}(t)\tilde{\Phi}_1(t)\}\\
			&-2\gamma_{\phi_2}^{-1}\int_{-\tau}^{0}\text{Tr}\{\dot{\Phi}_2^{*T}(t,\eta)\tilde{\Phi}_2(t,\eta)\}\text{d}\eta,
		\end{aligned}
	\end{equation}
	using $\text{Tr}(AB)=\text{Tr}(BA)$ and rearranging, we get
	\begin{equation} \label{p1}
		\begin{aligned}
			\dot{V}_2&=-e_y^T(t)Q_2e_y(t)+N^*(t)+\dot{W}(t)\\
			&+2\tilde{\lambda}_3^T(t)\left(-\text{diag}(L_r\Delta y(t-\tau))B_p^TP_2e_y(t)+\gamma_3^{-1}\dot{\hat{\lambda}}_3(t)\right)\\
			&+2\tilde{\lambda}_2^T(t)\Lambda_2(t)\left(\text{diag}(L_r\mathcal{G}(t-\tau))B_p^TP_2e_y(t)+\gamma_2^{-1}\dot{\hat{\lambda}}_2(t)\right)\\
			&+2\text{Tr}\{\tilde{\Phi}_1(t)\left(x_p(t-\tau)e_y^T(t)P_2B_pL_r+\gamma_{\phi_1}^{-1}\dot{\hat{\Phi}}_1^T(t)\right)\}\\
			&+2\int_{-\tau}^{0}\text{Tr}\{\tilde{\Phi}_2(t,\eta)\Big(L_ry_h(t+\eta-\tau)e_y^T(t)P_2B_pL_r+\gamma_{\phi_2}^{-1}\dot{\hat{\Phi}}_2^T(t,\eta)\Big)\}\text{d}\eta\\
			&-2\int_{-\tau}^{0}\dot{\tilde{\lambda}}_2^T(t+\nu)\Lambda_2(t)\text{diag}(L_r\mathcal{G}(t-\tau))B_p^TP_2e_y(t)\text{d}\nu\\
			&-2\int_{-\tau}^{0}\text{Tr}\{\dot{\tilde{\Phi}}_1(t+\nu)x_p(t-\tau)e_y^T(t)P_2B_pL_r\}\text{d}\nu\\
			&-2\int_{-\tau}^{0}\int_{-\tau}^{0}\text{Tr}\{\dot{\tilde{\Phi}}_2(t+\nu,\eta) L_ry_h(t+\eta-\tau)e_y^T(t)P_2B_pL_r\}\text{d}\nu\text{d}\eta.\\
		\end{aligned}
	\end{equation}
	Substituting the adaptive laws (\ref{outer update laws}) in (\ref{p1}) yields 
	\begin{equation} \label{p2}
		\begin{aligned}
			\dot{V}_2&=-e_y^T(t)Q_2e_y(t)+N^*(t)+\dot{W}(t)\\
			&+2\tilde{\lambda}_3^T(t)\left(\text{Proj}(\hat{\lambda}_3(t),\, Y_3(t))-Y_3(t)\right)\\
			&+2\tilde{\lambda}_2^T(t)\Lambda_2(t)\left(\text{Proj}(\hat{\lambda}_2(t),\, Y_2(t))-Y_2(t)\right)\\
			&+2\text{Tr}\{\tilde{\Phi}_1(t)\left(\text{Proj}(\hat{\Phi}_1(t),\, Y_{\phi_1}(t))-Y_{\phi_1}(t)\right)\}\\
			&+2\int_{-\tau}^{0}\text{Tr}\{\tilde{\Phi}_2(t,\eta)\Big(\text{Proj}(\hat{\Phi}_2(t,\eta),\, Y_{\phi_2}(t,\eta))-Y_{\phi_2}(t,\eta)\Big)\}\text{d}\eta\\
			&+2\int_{-\tau}^{0}\dot{\tilde{\lambda}}_2^T(t+\nu)\Lambda_2(t)Y_2(t)\text{d}\nu+2\int_{-\tau}^{0}\text{Tr}\{\dot{\tilde{\Phi}}_1(t+\nu)Y_{\phi_1}(t)\}\text{d}\nu\\
			&+2\int_{-\tau}^{0}\int_{-\tau}^{0}\text{Tr}\{\dot{\tilde{\Phi}}_2(t+\nu,\eta)Y_{\phi_2}(t,\eta)\}\text{d}\nu\text{d}\eta,
		\end{aligned}
	\end{equation}
	where 
	\begin{equation}\label{Ys}
		\begin{aligned}
			Y_2(t)&\triangleq-\text{diag}(L_r\mathcal{G}(t-\tau))B_p^TP_2e_y(t),\\
			Y_3(t)&\triangleq \text{diag}(L_r\Delta y(t-\tau))B_p^TP_2e_y(t),\\
			Y_{\phi_1}(t)&\triangleq -x_p(t-\tau)e_y^T(t)P_2B_pL_r,\\
			Y_{\phi_2}(t,\eta)&\triangleq -L_ry_h(t+\eta-\tau)e_y^T(t)P_2B_pL_r.
		\end{aligned}
	\end{equation}

	Using the projection property $(\theta_{i,j}-\theta^*_{i,j})(\mathrm{Proj}(\theta_{i,j},Y_{i,j})-Y_{i,j})\leq0$, and the fact that $\Lambda_2(t)$ is diagonal positive definite, it follows from (\ref{p2}) that
	\begin{equation*}
		\begin{aligned}
			\dot{V}_2&\leq-e_y^T(t)Q_2e_y(t)+N^*(t)+\dot{W}(t)\\
			&+2\int_{-\tau}^{0}\dot{\tilde{\lambda}}_2^T(t+\nu)\Lambda_2(t)Y_2(t)\text{d}\nu+2\int_{-\tau}^{0}\text{Tr}\{\dot{\tilde{\Phi}}_1(t+\nu)Y_{\phi_1}(t)\}\text{d}\nu\\
			&+2\int_{-\tau}^{0}\int_{-\tau}^{0}\text{Tr}\{\dot{\tilde{\Phi}}_2(t+\nu,\eta)Y_{\phi_2}(t,\eta)\}\text{d}\nu\text{d}\eta.
		\end{aligned}
	\end{equation*}
	Using the algebraic inequality $\text{Tr}\{2A^TB\}\leq\text{Tr}\{A^TA+B^TB\}$, we obtain that
	\begin{equation} \label{p3}
		\begin{aligned}
			\dot{V}_2&\leq-e_y^T(t)Q_2e_y(t)+N^*(t)+\dot{W}(t)\\
			&+\int_{-\tau}^{0}\dot{\tilde{\lambda}}_2^T(t+\nu)\dot{\tilde{\lambda}}_2(t+\nu)\text{d}\nu+\int_{-\tau}^{0}Y_2^T(t)\Lambda_2(t)\Lambda_2(t)Y_2(t)\text{d}\nu\\
			&+\int_{-\tau}^{0}\text{Tr}\{\dot{\tilde{\Phi}}^T_1(t+\nu)\dot{\tilde{\Phi}}_1(t+\nu)\}\text{d}\nu+\int_{-\tau}^{0}\text{Tr}\{Y_{\phi_1}^T(t)Y_{\phi_1}(t)\}\text{d}\nu\\
			&+\int_{-\tau}^{0}\int_{-\tau}^{0}\text{Tr}\{\dot{\tilde{\Phi}}^T_2(t+\nu,\eta)\dot{\tilde{\Phi}}_2(t+\nu,\eta)\}\text{d}\nu\text{d}\eta\\
			&+\int_{-\tau}^{0}\int_{-\tau}^{0}\text{Tr}\{Y_{\phi_2}^T(t,\eta)Y_{\phi_2}(t,\eta)\}\text{d}\nu\text{d}\eta.
		\end{aligned}
	\end{equation}
	Substituting (\ref{Wdot}) in (\ref{p3}) yields
	\begin{equation}\label{p4}
		\begin{aligned}
			\dot{V}_2&\leq-e_y^T(t)Q_2e_y(t)+N^*(t)\\
			&+\tau Y_2^T(t)\Lambda_2(t)\Lambda_2(t)Y_2(t)+\tau\text{Tr}\{Y_{\phi_1}^T(t)Y_{\phi_1}(t)\}\\
			&+\tau\int_{-\tau}^{0}\text{Tr}\{Y_{\phi_2}^T(t,\eta)Y_{\phi_2}(t,\eta)\}\text{d}\eta+\tau\dot{\tilde{\lambda}}_2^T(t)\dot{\tilde{\lambda}}_2(t)\\
			&+\tau\text{Tr}\{\dot{\tilde{\Phi}}_1^T(t)\dot{\tilde{\Phi}}_1(t)\}+\tau\int_{-\tau}^{0}\text{Tr}\{\dot{\tilde{\Phi}}_2^T(t,\eta)\dot{\tilde{\Phi}}_2(t,\eta)\}\text{d}\eta.
		\end{aligned}
	\end{equation}
	Using the algebraic inequality 
	\begin{equation*}
		\begin{aligned}
			\text{Tr}\{\dot{\tilde{\Phi}}^T_1(t)\dot{\tilde{\Phi}}_1(t)\}&=\text{Tr}\{\dot{\hat{\Phi}}_1^T(t)\dot{\hat{\Phi}}_1(t)\}+\text{Tr}\{\dot{\Phi}_1^{*T}(t)\dot{\Phi}^*_1(t)\}-2\text{Tr}\{\dot{\hat{\Phi}}^T(t)\dot{\Phi}^*(t)\}\\
			&\leq 2\text{Tr}\{\dot{\hat{\Phi}}_1^T(t)\dot{\hat{\Phi}}_1(t)\}+2\text{Tr}\{\dot{\Phi}_1^{*T}(t)\dot{\Phi}^*_1(t)\},
		\end{aligned}
	\end{equation*}
	for the last three terms in (\ref{p4}), once can write
	\begin{equation}\label{p5}
		\begin{aligned}
			\dot{V}_2&\leq-e_y^T(t)Q_2e_y(t)+N^*(t)\\
			&+\tau Y_2^T(t)\Lambda_2(t)\Lambda_2(t)Y_2(t)+\tau\text{Tr}\{Y_{\phi_1}^T(t)Y_{\phi_1}(t)\}\\
			&+\tau\int_{-\tau}^{0}\text{Tr}\{Y_{\phi_2}^T(t,\eta)Y_{\phi_2}(t,\eta)\}\text{d}\eta\\
			&+2\tau\dot{\hat{\lambda}}_2^T(t)\dot{\hat{\lambda}}_2(t)+2\tau\dot{\lambda}_2^{*T}(t)\dot{\lambda}^*_2(t)\\
			&+2\tau\text{Tr}\{\dot{\hat{\Phi}}_1^T(t)\dot{\hat{\Phi}}_1(t)\}+2\tau\text{Tr}\{\dot{\Phi}_1^{*T}(t)\dot{\Phi}^*_1(t)\}\\
			&+2\tau\int_{-\tau}^{0}\text{Tr}\{\dot{\hat{\Phi}}_2^T(t,\eta)\dot{\hat{\Phi}}_2(t,\eta)\}\text{d}\eta+2\tau\int_{-\tau}^{0}\text{Tr}\{\dot{\Phi}_2^{*T}(t,\eta)\dot{\Phi}^*_2(t,\eta)\}\text{d}\eta.
		\end{aligned}
	\end{equation}
	Let $Y_2(t)=[a_1(t),\dots,a_m(t)]^T$ and $\Lambda_2(t)=\text{diag}([b_1(t),\dots,b_m(t)])$ for some $a_i(t)\in\mathbb{R},\,b_i(t)\in \mathbb{R}_+$, for $i=1,\dots,m$. Since $\Lambda_2(t)$, defined in (\ref{Lambda_2(t)}), is shown to be bounded in Remark 1, then $b_i(t)\leq\lambda_{i,i}\hat{\lambda}_{max_i}$ for all $t\geq t_0$, $i=1,\dots,m$, where $\lambda_{i,i}$ is the $i^{th}$ diagonal element of $\Lambda$, and $\hat{\lambda}_{max_i}$ is the $i^{th}$ projection upper bound of $\hat{\lambda}(t)$ in (\ref{lmbda}). Then, one can write
	\begin{equation}\label{p6}
		\begin{aligned}
			Y_2^T(t)\Lambda_2(t)\Lambda_2(t)Y_2(t)=&b_1^2(t)a_1^2(t)+\dots+b_m^2(t)a_m^2(t)\\
			&\leq\mu(a_1^2(t)+\dots+a_m^2(t))=\mu Y_2^T(t)Y_2(t),
		\end{aligned}
	\end{equation}
	where $\mu\triangleq \max_i(\lambda_{i,i}\hat{\lambda}_{max_i})^2$.
	Furthermore, using the property that the projection operator bounds an adaptive parameter in a compact set, then from the element-wise projection operator's definition in \cite{ShahabAuto}, it can be shown that
	\begin{equation}\label{p7}
		\text{Tr}\{\dot{\hat{\Phi}}_1^T(t)\dot{\hat{\Phi}}_1(t)\}\leq\gamma_{\phi_1}^2\text{Tr}\{Y_{\phi_1}^T(t)Y_{\phi_1}(t)\}.
	\end{equation}
	Using (\ref{p6}) and (\ref{p7}) in (\ref{p5}), we obtain that
	\begin{equation}\label{p8}
		\begin{aligned}
			\dot{V}_2&\leq-e_y^T(t)Q_2e_y(t)+N^*(t)\\
			&+\tau\mu Y_2^T(t)Y_2(t)+\tau\text{Tr}\{Y_{\phi_1}^T(t)Y_{\phi_1}(t)\}+\tau\int_{-\tau}^{0}\text{Tr}\{Y_{\phi_2}^T(t,\eta)Y_{\phi_2}(t,\eta)\}\text{d}\eta\\
			&+2\tau \gamma_2^2Y_2^T(t)Y_2(t)+2\tau\dot{\lambda}_2^{*T}(t)\dot{\lambda}^*_2(t)\\
			&+2\tau\gamma_{\phi_1}^2\text{Tr}\{Y_{\phi_1}^T(t)Y_{\phi_1}(t)\}+2\tau\text{Tr}\{\dot{\Phi}_1^{*T}(t)\dot{\Phi}^*_1(t)\}\\
			&+2\tau\gamma_{\phi_2}^2\int_{-\tau}^{0}\text{Tr}\{Y_{\phi_2}^T(t,\eta)Y_{\phi_2}(t,\eta)\}\text{d}\eta+2\tau\int_{-\tau}^{0}\text{Tr}\{\dot{\Phi}_2^{*T}(t,\eta)\dot{\Phi}^*_2(t,\eta)\}\text{d}\eta\\
			&=-e_y^T(t)Q_2e_y(t)+N^*(t)\\
			&+\tau(\mu+2\gamma_2^2) Y_2^T(t)Y_2(t)+\tau(1+2\gamma_{\phi_1}^2)\text{Tr}\{Y_{\phi_1}^T(t)Y_{\phi_1}(t)\}\\
			&+\tau(1+2\gamma_{\phi_2}^2)\int_{-\tau}^{0}\text{Tr}\{Y_{\phi_2}^T(t,\eta)Y_{\phi_2}(t,\eta)\}\text{d}\eta\\
			&+2\tau\dot{\lambda}_2^{*T}(t)\dot{\lambda}^*_2(t)+2\tau\text{Tr}\{\dot{\Phi}_1^{*T}(t)\dot{\Phi}^*_1(t)\}\\
			&+2\tau\int_{-\tau}^{0}\text{Tr}\{\dot{\Phi}_2^{*T}(t,\eta)\dot{\Phi}^*_2(t,\eta)\}\text{d}\eta.\\
		\end{aligned}
	\end{equation}
	Using the property $\text{Tr}\{Y^TY\}=\norm{Y}_F^2$ for a matrix $Y$, we can rewrite (\ref{p8}) as
	\begin{equation}\label{p9}
		\begin{aligned}
			\dot{V}_2&\leq-e_y^T(t)Q_2e_y(t)+N^*(t)\\
			&+\tau(\mu+2\gamma_2^2) \norm{Y_2(t)}^2+\tau(1+2\gamma_{\phi_1}^2) \norm{Y_{\phi_1}(t)}_F^2\\
			&+\tau(1+2\gamma_{\phi_2}^2)\int_{-\tau}^{0}\norm{Y_{\phi_2}(t,\eta)}_F^2\text{d}\eta\\
			&+2\tau\dot{\lambda}_2^{*T}(t)\dot{\lambda}^*_2(t)+2\tau\text{Tr}\{\dot{\Phi}_1^{*T}(t)\dot{\Phi}^*_1(t)\}\\
			&+2\tau\int_{-\tau}^{0}\text{Tr}\{\dot{\Phi}_2^{*T}(t,\eta)\dot{\Phi}^*_2(t,\eta)\}\text{d}\eta.\\
		\end{aligned}
	\end{equation}
	Substituting (\ref{Ys}) into (\ref{p9}), yields
	\begin{equation}\label{p10}
		\begin{aligned}
			\dot{V}_2&\leq-e_y^T(t)Q_2e_y(t)+N^*(t)\\
			&+\tau(\mu+2\gamma_2^2) \norm{\text{diag}(L_r\mathcal{G}(t-\tau))B_p^TP_2e_y(t)}^2\\
			&+\tau(1+2\gamma_{\phi_1}^2) \norm{x_p(t-\tau)e_y^T(t)P_2B_pL_r}_F^2\\
			&+\tau(1+2\gamma_{\phi_2}^2)\int_{-\tau}^{0}\norm{L_ry_h(t+\eta-\tau)e_y^T(t)P_2B_pL_r}_F^2\text{d}\eta\\
			&+2\tau\dot{\lambda}_2^{*T}(t)\dot{\lambda}^*_2(t)+2\tau\text{Tr}\{\dot{\Phi}_1^{*T}(t)\dot{\Phi}^*_1(t)\}\\
			&+2\tau\int_{-\tau}^{0}\text{Tr}\{\dot{\Phi}_2^{*T}(t,\eta)\dot{\Phi}^*_2(t,\eta)\}\text{d}\eta\\
			&\leq -\lambda_{min}(Q_2)\norm{e_y(t)}^2+N^*(t)\\
			&+\tau(\mu+2\gamma_2^2) \norm{\text{diag}(L_r\mathcal{G}(t-\tau))}^2\norm{P_2B_p}^2\norm{e_y(t)}^2\\
			&+\tau(1+2\gamma_{\phi_1}^2) \norm{x_p(t-\tau)}^2\norm{e_y(t)}^2\norm{P_2B_pL_r}_F^2\\
			&+\tau(1+2\gamma_{\phi_2}^2)\int_{-\tau}^{0}\norm{L_ry_h(t+\eta-\tau)}^2\norm{e_y(t)}^2\norm{P_2B_pL_r}_F^2\text{d}\eta\\
			&+2\tau\dot{\lambda}_2^{*T}(t)\dot{\lambda}^*_2(t)+2\tau\text{Tr}\{\dot{\Phi}_1^{*T}(t)\dot{\Phi}^*_1(t)\}\\
			&+2\tau\int_{-\tau}^{0}\text{Tr}\{\dot{\Phi}_2^{*T}(t,\eta)\dot{\Phi}^*_2(t,\eta)\}\text{d}\eta.\\
		\end{aligned}
	\end{equation}
	Substituting (\ref{N*}) for $N^*(t)$ in (\ref{p10}), yields
	\begin{equation*}
		\begin{aligned}
			\dot{V}_2&\leq -\lambda_{min}(Q_2)\norm{e_y(t)}^2\\
			&+\tau(\mu+2\gamma_2^2) \norm{\text{diag}(L_r\mathcal{G}(t-\tau))}^2\norm{P_2B_p}^2\norm{e_y(t)}^2\\
			&+\tau(1+2\gamma_{\phi_1}^2) \norm{x_p(t-\tau)}^2\norm{e_y(t)}^2\norm{P_2B_pL_r}_F^2\\
			&+\tau(1+2\gamma_{\phi_2}^2)\int_{-\tau}^{0}\norm{L_ry_h(t+\eta-\tau)}^2\norm{e_y(t)}^2\norm{P_2B_pL_r}_F^2\text{d}\eta\\
			&+2\tau\dot{\lambda}_2^{*T}(t)\dot{\lambda}^*_2(t)+2\tau\text{Tr}\{\dot{\Phi}_1^{*T}(t)\dot{\Phi}^*_1(t)\}\\
			&+2\tau\int_{-\tau}^{0}\text{Tr}\{\dot{\Phi}_2^{*T}(t,\eta)\dot{\Phi}^*_2(t,\eta)\}\text{d}\eta\\
			&+\gamma_2^{-1}\tilde{\lambda}_2^T(t)\dot{\Lambda}_2(t)\tilde{\lambda}_2(t)-2\gamma_2^{-1}\tilde{\lambda}_2^T(t)\Lambda_2(t)\dot{\lambda}^*_2(t)\\
			&-2\gamma_3^{-1}\tilde{\lambda}_3^T(t)\dot{\lambda}^*_3(t)-2\gamma_{\phi_1}^{-1}\text{Tr}\{\dot{\Phi}_1^{*T}(t)\tilde{\Phi}_1(t)\}\\
			&-2\gamma_{\phi_2}^{-1}\int_{-\tau}^{0}\text{Tr}\{\dot{\Phi}_2^{*T}(t,\eta)\tilde{\Phi}_2(t,\eta)\}\text{d}\eta.
		\end{aligned}
	\end{equation*}
	Using Remarks 1 and 2, and denoting $p\triangleq \max(\norm{P_2B_p}^2,\norm{P_2B_pL_r}_F^2)$, yields
	\begin{equation}\label{p11}
		\begin{aligned}
			\dot{V}_2&\leq -\lambda_{min}(Q_2)\norm{e_y(t)}^2\\
			&+\tau p(\mu+2\gamma_2^2) \norm{\text{diag}(L_r\mathcal{G}(t-\tau))}^2\norm{e_y(t)}^2\\
			&+\tau p(1+2\gamma_{\phi_1}^2)\norm{x_p(t-\tau)}^2\norm{e_y(t)}^2\\
			&+\tau p(1+2\gamma_{\phi_2}^2)\int_{-\tau}^{0}\norm{L_ry_h(t+\eta-\tau)}^2\norm{e_y(t)}^2\text{d}\eta\\
			&+2\tau\dot{\beta}_2^2+2\tau\dot{\phi}_1^2+2\tau^2\dot{\phi}_2^2+\gamma_2^{-1}\tilde{\beta}_2^2\dot{\beta}_3+2\gamma_2^{-1}\tilde{\beta}_2\beta_3\dot{\beta}_2\\
			&+2\gamma_3^{-1}\tilde{\beta}_3\dot{\beta}_3+2\gamma_{\phi_1}^{-1}\dot{\phi}_1\tilde{\phi}_1+2\gamma_{\phi_2}^{-1}\tau\dot{\phi}_2\tilde{\phi}_2,
		\end{aligned}
	\end{equation}
	where $\tilde{\beta}_2\triangleq \norm{\hat{\lambda}_{2max}}+\beta_2$, $\tilde{\beta}_3\triangleq \norm{\hat{\lambda}_{3max}}+\beta_3$, $\tilde{\phi}_1\triangleq \norm{\hat{\Phi}_{1max}}+\phi_1$, and $\tilde{\phi}_2\triangleq \norm{\hat{\Phi}_{2max}}+\phi_2$. Defining $q\triangleq \lambda_{min}(Q_2)/p$, and rearranging, one can rewrite (\ref{p11}) as
	\begin{equation}\label{p12}
		\begin{aligned}
			\dot{V}_2&\leq \, p\norm{e_y(t)}^2\Big(-q+\tau\Big\{  (\mu+2\gamma_2^2) \norm{\text{diag}(L_r\mathcal{G}(t-\tau))}^2+(1+2\gamma_{\phi_1}^2)\norm{x_p(t-\tau)}^2\\
			&\:\:\:\:\:\:\:\:\:\:\:\:\:\:\:\:\:\:\:\:\:\:\:\:\:\:\:\:\:+(1+2\gamma_{\phi_2}^2)\int_{-\tau}^{0}\norm{L_ry_h(t+\eta-\tau)}^2\text{d}\eta\Big\} \Big)\\
			&+2\tau(\dot{\beta}_2^2+\dot{\phi}_1^2+\tau\dot{\phi}_2^2)+\gamma_2^{-1}\tilde{\beta}_2^2\dot{\beta}_3+2\gamma_2^{-1}\tilde{\beta}_2\beta_3\dot{\beta}_2\\
			&+2\gamma_3^{-1}\tilde{\beta}_3\dot{\beta}_3+2\gamma_{\phi_1}^{-1}\dot{\phi}_1\tilde{\phi}_1+2\gamma_{\phi_2}^{-1}\tau\dot{\phi}_2\tilde{\phi}_2.
		\end{aligned}
	\end{equation}
	It follows from Lemma 1 that $x_p(t)$ is bounded, which implies that there exists $\alpha_1\in \mathbb{R}_+$ such that $\norm{x_p(t)}^2\leq \alpha_1$ for all $t\geq t_0$. Since $\norm{y_h(t)}\leq \norm{y_o}$ for all $t\geq t_0$ due to human input saturation, then there exists $\alpha_2\in\mathbb{R}_+$ such that $\norm{L_ry_h(t)}^2\leq \alpha_2$ for all $t\geq t_0$. In addition, as $r(t)$ is bounded, then all the terms in (\ref{G}) are bounded due to the usage of the projection operator in (\ref{outer update laws}), which implies the boundedness of $\mathcal{G}(t)$ and hence the existence of $\alpha_3\in \mathbb{R}_+$ such that $\norm{\text{diag}(L_r\mathcal{G}(t))}^2\leq \alpha_3$ for all $t\geq t_0$. Therefore, using (\ref{p12}), one can write 
	\begin{equation}\label{p13}
		\begin{aligned}
			\dot{V}_2&\leq \, p\norm{e_y(t)}^2\Big(-q+\tau\Big\{  (\mu+2\gamma_2^2) \alpha_3+(1+2\gamma_{\phi_1}^2)\alpha_1+(1+2\gamma_{\phi_2}^2)\tau \alpha_2\Big\} \Big)\\
			&+2\tau(\dot{\beta}_2^2+\dot{\phi}_1^2+\tau\dot{\phi}_2^2)+\gamma_2^{-1}\tilde{\beta}_2^2\dot{\beta}_3+2\gamma_2^{-1}\tilde{\beta}_2\beta_3\dot{\beta}_2\\
			&+2\gamma_3^{-1}\tilde{\beta}_3\dot{\beta}_3+2\gamma_{\phi_1}^{-1}\dot{\phi}_1\tilde{\phi}_1+2\gamma_{\phi_2}^{-1}\tau\dot{\phi}_2\tilde{\phi}_2.
		\end{aligned}
	\end{equation}
	Then, there exists a small enough $\tau^*\in \mathbb{R}_+$ such that 
	\begin{equation}\label{tau*}
		\tau^*\Big\{  (\mu+2\gamma_2^2) \alpha_3+(1+2\gamma_{\phi_1}^2)\alpha_1+(1+2\gamma_{\phi_2}^2)\tau^* \alpha_2\Big\}<q,
	\end{equation}
	which implies, from (\ref{p13}), that for any $\tau\in[0,\tau^*]$, $\dot{V}_2<0$ whenever 
	\begin{equation}
		\norm{e_y(t)}^2>\frac{z_1}{z_2},
	\end{equation}
	where
	\begin{equation}\label{z_1}
		\begin{aligned}
			z_1\triangleq&2\tau(\dot{\beta}_2^2+\dot{\phi}_1^2+\tau\dot{\phi}_2^2)+\gamma_2^{-1}\tilde{\beta}_2^2\dot{\beta}_3+2\gamma_2^{-1}\tilde{\beta}_2\beta_3\dot{\beta}_2+2\gamma_3^{-1}\tilde{\beta}_3\dot{\beta}_3+2\gamma_{\phi_1}^{-1}\dot{\phi}_1\tilde{\phi}_1+2\gamma_{\phi_2}^{-1}\tau\dot{\phi}_2\tilde{\phi}_2, 
		\end{aligned}
	\end{equation}
	\begin{equation}\label{z_2}
		\begin{aligned}
			z_2\triangleq p\Big(& q-\tau\Big\{  (\mu+2\gamma_2^2) \alpha_3+(1+2\gamma_{\phi_1}^2)\alpha_1+(1+2\gamma_{\phi_2}^2)\tau \alpha_2\Big\}\Big).
		\end{aligned}
	\end{equation}
	Hence, for any $\tau\in[0,\tau^*]$, the solution $(e_y(t), \tilde{\lambda}_2(t),\tilde{\lambda}_3(t),\tilde{\Phi}_1(t), \tilde{\Phi}_2(t,\eta))$ is bounded and converges to the compact set
	\begin{equation}\label{E}
		\begin{aligned}
			E\triangleq \bigg\{&(e_y(t), \tilde{\lambda}_2(t),\tilde{\lambda}_3(t),\tilde{\Phi}_1(t), \tilde{\Phi}_2(t,\eta)):\norm{e_y(t)}^2\leq\frac{z_1}{z_2},\\
			&\norm{\tilde{\lambda}_2(t)}\leq\tilde{\beta}_2, \norm{\tilde{\lambda}_3(t)}\leq \tilde{\beta}_3, \norm{\tilde{\Phi}_1(t)}\leq \tilde{\phi}_1, \norm{\tilde{\Phi}_2(t,\eta)}\leq\tilde{\phi}_2\bigg\}.
		\end{aligned}
	\end{equation}
	According to (\ref{crossover refernece model}), since $r(t)$ is bounded, then so is $x_m(t)$. This with the fact that $x_p(t)$ is bounded (by Lemma 1) imply the boundedness of $e_2(t)$. And since $e_y(t)=e_2(t)-e_{\Delta}(t)$ is bounded, then $e_{\Delta}(t)$ is also bounded, which implies by (\ref{auxilliary error}) that $\Delta y(t)$ is bounded and completes the proof.\hfill $\blacksquare$
	
	\textbf{Remark 3:} While the existence of the upper bound on the time delay $\tau^*$ is guaranteed, its value depends on the selection of the outer-loop learning rates $\gamma_{out}\triangleq\{\gamma_2,\gamma_{\phi_1},\gamma_{\phi_2}\}$. Note from (\ref{tau*}) that as larger values of $\gamma_{out}$ are used, the allowable maximum time delay $\tau^*$ becomes smaller. On the other hand, in the limit where $\gamma_{out}\to0$, which corresponds to no adaptation, $\tau^*$ approaches its ultimate value $\tau_{max}$ satisfying
	\begin{equation}\label{tau_max}
		\tau_{max}\Big(  \mu \alpha_3+\alpha_1+\tau_{max} \alpha_2\Big)<q.
	\end{equation}
	On the contrary, the ultimate bound $z\triangleq z_1/z_2$ on the error $e_y(t)$, which is defined by the set (\ref{E}), (\ref{z_1}) and (\ref{z_2}), is inversely proportional to the values of $\gamma_{out}$. That is, to achieve a better tracking performance, which corresponds to smaller values of $z$, the outer-loop learning rates $\gamma_{out}$ should be selected as large as possible. And in the limit where $\gamma_{out}\to \infty$, the upper bound $z\to 0$. Therefore, given any delay value $\tau<\tau_{max}$, the optimal outer-loop learning rates $\gamma_{out,opt}$ are the ones that satisfy $\tau^*=\tau$. A further increase in $\gamma_{out}>\gamma_{out,opt}$ renders our stability analysis inapplicable due to $\tau>\tau^*$, while a decrease in $\gamma_{out}<\gamma_{out,opt}$ allows for higher delay values to be tolerated at the expense of a deteriorated tracking performance.
	
	%%%%%%%%%%%%%%%%%%%%%%%%%%%%%%%%%%%%%%
	\section{Simulations} \label{Simulations}
	Consider the perturbation equations of the longitudinal motion for the 747 airplane \cite{bryson} cruising in level flight at an altitude of 40 kft and a velocity of 774 ft/sec with the dynamics given in the form of (\ref{plant dynamics}). The state vector is
	\begin{equation}
		x_p(t)=
		\begin{bmatrix}
			x_{p1}(t) & x_{p2}(t) & x_{p3}(t) & x_{p4}(t)
		\end{bmatrix}^T,
	\end{equation}
	where $x_{p1}(t)$ and $x_{p2}(t)$ are the components of the aircraft's velocity along the $x$ and $z$-axes, respectively, with respect to the reference axis (in ft/sec), $x_{p3}(t)$ is the aircraft's pitch rate (in crad/sec), and $x_{p4}(t)$ is the pitch angle of the aircraft (in crad). The input $u_p(t)$ represents the elevator deflection (in crad), and the nominal system and control input matrices are given by
	\begin{equation}\label{nomnom}
		\begin{aligned}
			A_n=&
			\begin{bmatrix}
				-0.0030 & 0.0390 & 0 & -0.3220 \\
				-0.0650 & -0.3190 & 7.7400 & 0 \\
				0.0200 & -0.1010 & -0.4290 & 0 \\
				0 & 0 & 1 & 0 
			\end{bmatrix},\\
			B_p=&
			\begin{bmatrix}
				0.0100 &
				-0.1800 &
				-1.1600&
				0 
			\end{bmatrix}^T,
		\end{aligned}
	\end{equation}
	with the eigenvalues at $-0.3750\pm0.8818i$ and $-0.0005\pm0.0674i$. We consider an uncertainty in the system matrix $A_p$ constructed as
	\begin{equation}
		A_p=
		\begin{bmatrix}
			-0.0029 & 0.0389 & -0.0047 & -0.3220 \\
			-0.0661 & -0.3171 & 7.8254 & 0.0008 \\
			0.0129 & -0.0888 & 0.1210 & 0.0051\\
			0 & 0 & 1 & 0 
		\end{bmatrix},
	\end{equation}
	such that the eigenvalues are placed at 0.1, 0.2, 0.3 and 0.4 in the right-half complex plane. A pilot is controlling the aircraft to achieve a desired pitch angle by feeding pitch rate commands to the inner-loop controller, i.e., $y_1(t)=x_{p3}$, and $y_2(t)=x_{p4}(t)$, where the pilot command is saturated as in (\ref{saturation}), with $y_o=10$ deg/s.
	
	The reference model dynamics are assigned in the form of (\ref{reference model}), where $A_r=A_n$ and $B_r=B_pL_r$. Achieving pilot command following by assigning the feed-forward gain $L_r$ as in (\ref{Lr}) is not possible since the transfer function $x_{p3}(s)/u_p(s)$ has a zero at the origin. Instead, we design the inner-loop feed-forward controller by assuming short-period dynamics approximation of the nominal dynamics given by
	\begin{equation}
		A_{sp}=
		\begin{bmatrix}
			-0.3190 & 7.7400 \\
			-0.1010 & -0.4290 \\
		\end{bmatrix},
		\;\;
		B_{sp}=
		\begin{bmatrix}
			-0.1800 \\
			-1.1600\\
		\end{bmatrix}.
	\end{equation} 
	
	The eigen values are at $-0.3740\pm 0.8824i$ which makes a good approximation of the fast dynamics (eigen values) of (\ref{nomnom}). Then, the feed-forward gain is selected as $L_r=-(C_{sp}^TA_{sp}^{-1}B_{sp})^{-1}$, where $C_{sp}=[0,\,1]^T$. For the outer-loop human pilot model, the LQR method is used to design the crossover-reference model (\ref{crossover refernece model}) by calculating $\theta_x$ using $Q_{LQR}=\text{diag}([0,\,0,\,0,\,3])$ and $R_{LQR}=3$, with $\theta_r$ assigned as in (\ref{thr}). We use $\tau=0.3$ s for the human internal time delay, which is determined by averaging the operators' delay in an adaptive pilot experiment \cite{Shahabpilot}. The Lyapunov matrices are taken as $Q_1=Q_2=0.001I_{4\times4}$. The finite integral term in (\ref{G}) and the adaptive law (\ref{phi2dot}) are implemented by discretizing the integral into 5 intervals as illustrated in \cite{yildizspark}. The adaptive parameters $\hat{\lambda}(t)$, $\hat{\lambda}_2(t)$ and $\hat{\lambda}_3(t)$ are initialized at 1, $\hat{\Phi}_1(t)$ is initialized at $-\theta_x e^{A_r\tau}$, and the rest are initialized at zero. Finally, the outer-loop learning rates are taken as $\gamma_2=1$, $\gamma_3=5$, $\gamma_{\phi_1}=\text{diag}([0.01,\,0.001,\,0.01,\,0.01])$ and $\gamma_{\phi_2}=0.1$, and the inner loop learning rate is fixed at $\gamma_\lambda=1$.
	
	\begin{figure}[t]
		\centering
		\includegraphics[width=\linewidth]{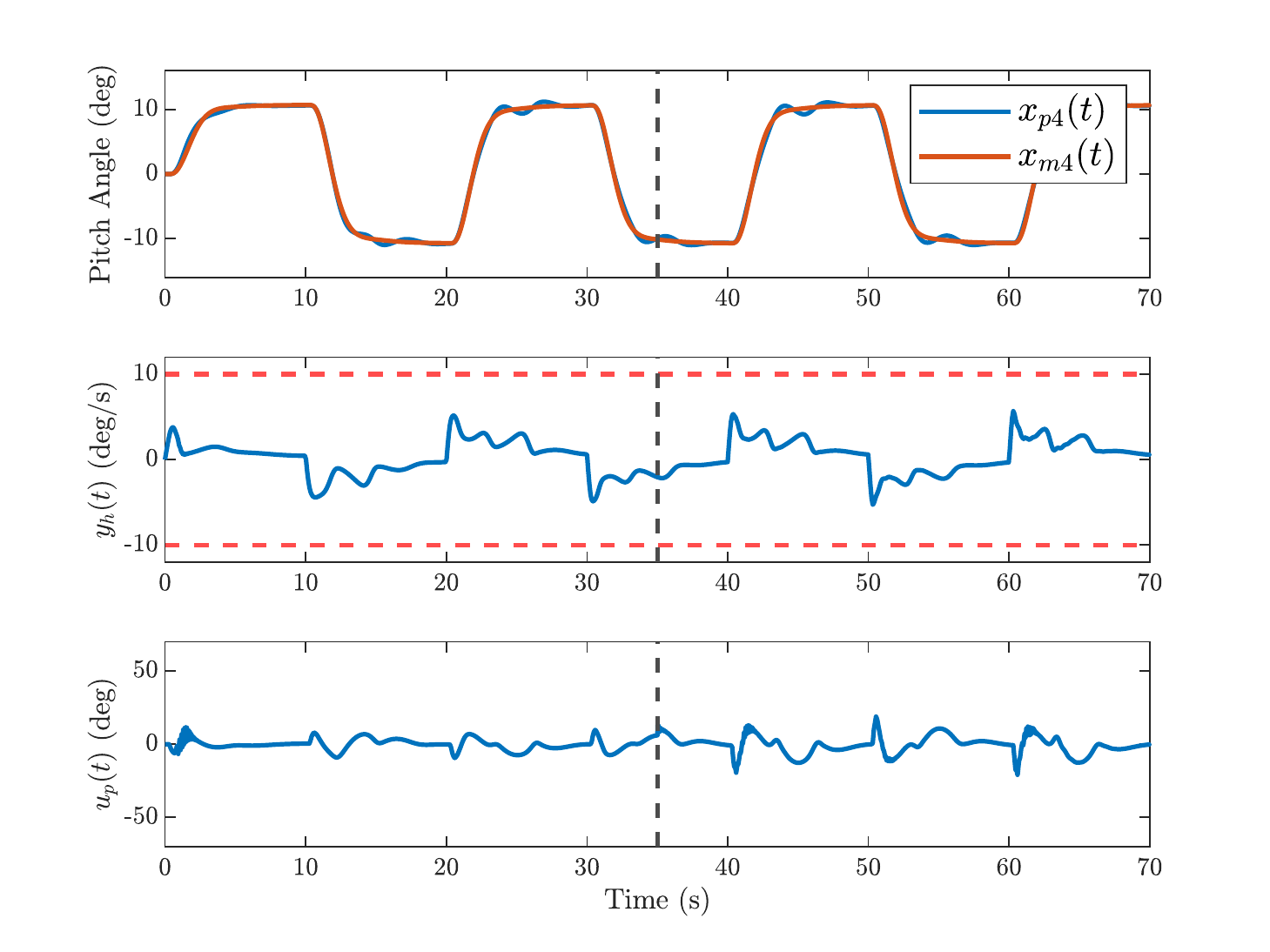}
		\caption{Pitch angle, pilot commands and controller input for $\gamma_x=1$.}
		\label{gx1}
	\end{figure}
	\begin{figure}[t]
		\centering
		\includegraphics[width=\linewidth]{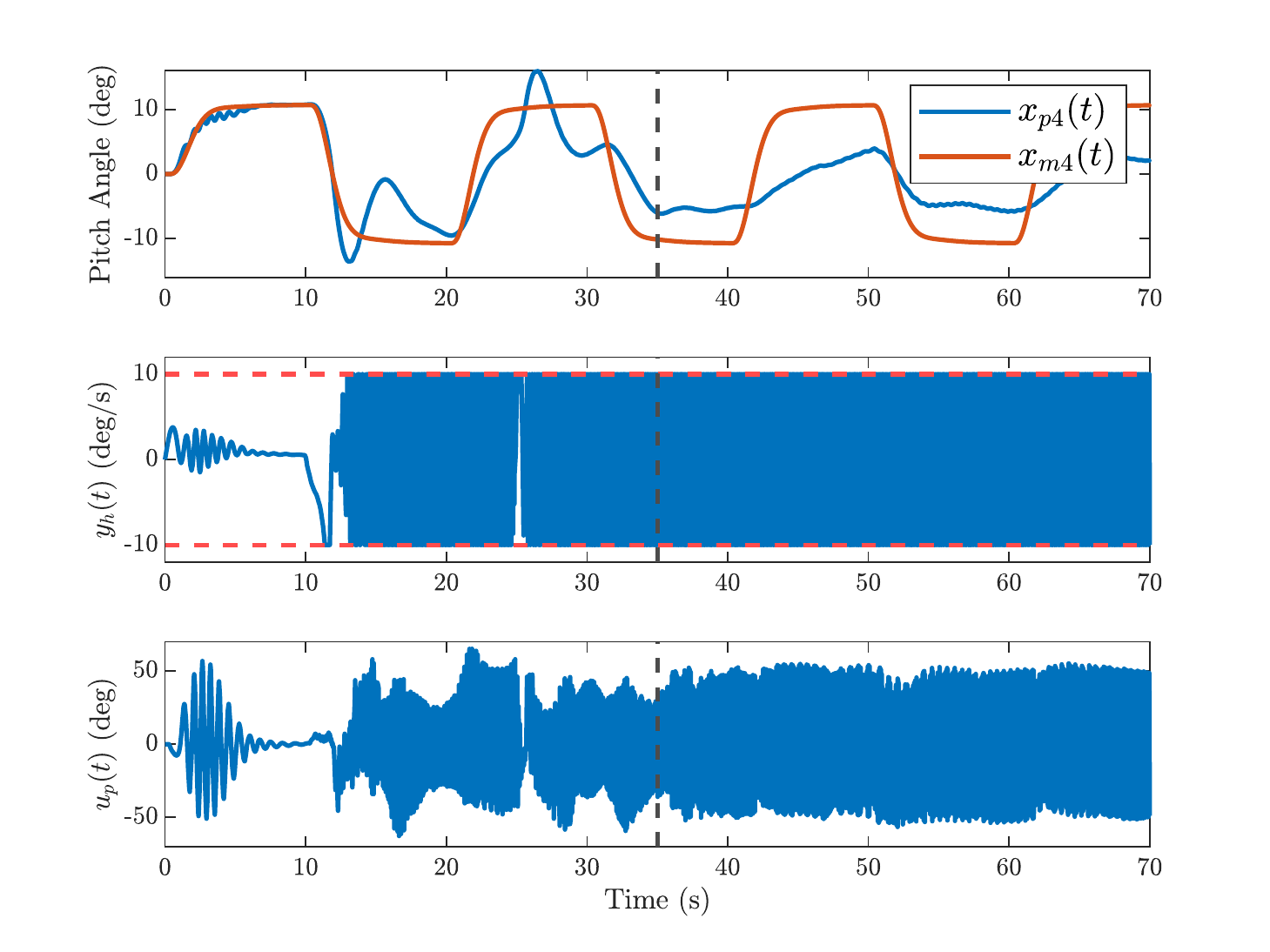}
		\caption{Pitch angle, pilot commands and controller input for $\gamma_x=0.01$.}
		\label{gx0.01}
	\end{figure}
	
	Figs. \ref{gx1} and \ref{gx0.01} show the aircraft pitch angle, the evolution of the pilot commands and the plant control input for different inner-loop learning rates. We start with $\Lambda=1$, and we introduce a failure into the system by making $\Lambda=0.6$ for $t\geq35$ s. Good tracking performance is achieved with a reasonable control effort of both the pilot and the controller in Fig. \ref{gx1}, where $\gamma_x=1$. As the inner-loop learning rate is decreased to $\gamma_x=0.01$ in Fig. \ref{gx0.01}, a significant deterioration in the tracking performance is observed, accompanied by saturating high frequency oscillations in the pilot commands and the controller input. This showcases an example of a poor adaptive controller design, where the pilot spends a large control effort to maintain a satisfactory performance.

	%%%%%%%%%%%%%%%%%%%%%%%%%%%%%%%%%%%%%%%%%%%%%%%%%%%%%%%%%%%%%%%%%%%%%%%%%%%%%%%%
	%%%%%%%%%%%%%%%%%%%%%%%%%%%%%%%%%%%%%%%%%%%%%%%%%%%%%%%%%%%%%%%%%%%%%%%%%%%%%%%%

	\bibliographystyle{unsrt}
	\bibliography{main.bib}

\end{document}